# Estimands and the Choice of Non-Inferiority Margin under ICH E9(R1)

**Authors:** Tobias Mütze[1], Helle Lynggaard[2], Sunita Rehal[3], Oliver N. Keene[4], Marian Mitroiu[5], David Wright[6]

**Affiliations:**

[1] Statistical Methodology, Novartis Pharma AG, Basel, Switzerland

[2] Biostatistics Methods, Bagsværd, Novo Nordisk A/S, Denmark

[3] GlaxoSmithKline Plc, London, UK, ORCID: 0000-0002-2578-5778

[4] KeeneONStatistics, Maidenhead, UK

[5] Biostatistics, Statistical Sciences and Evidence Generation, Biogen International GmbH, Baar, Switzerland

[6] Statistical Innovation, Respiratory and Immunology Biometrics and Statistical Innovation, Biopharmaceuticals R&D, AstraZeneca, Cambridge, UK

**Abstract:**

Since the release of the ICH E9(R1) addendum on estimands, its application in non-inferiority trials has received far less attention than in superiority settings. A key conclusion from Lynggaard et al. was that the "choice of non-inferiority margin must reflect the chosen estimand." However, current regulatory guidance predates ICH E9(R1) and therefore does not reflect how the estimand influences the historical evidence and constancy assumption (assay sensitivity) used to derive the non-inferiority margin. This paper investigates the degree to which the non-inferiority margin depends on the estimand.

Using simulated patient journeys in a weight-management setting, we illustrate how different intercurrent event strategies and variations in the intercurrent event frequency affect the estimand, and consequently the estimated treatment effect. These results emphasize that the historical treatment effect of the reference treatment versus placebo, and thus the margin $M_1$, is specific to an estimand and may differ even when trials formally target similar questions.

We further illustrate the process of determining the non-inferiority margin using two examples in non-inferiority trials for a new theoretical weight management treatment. In the first example, we focus on the setting where the historical clinical trials use the estimand framework highlighting that even when they include the estimand framework, determining the non-inferiority margin can be challenging in case the historical trials target an estimand different from the one in the planned study. A second example highlights challenges when historical trials did not employ the estimand framework and the targeted estimand cannot be fully reconstructed.

# 1 Introduction

Since the release of the ICH E9(R1) addendum on estimands and sensitivity analysis in clinical trials (1), substantial research has been published on its application in superiority trials (2–5), but its use in non-inferiority trials has received far less attention.

Lynggaard *et al.* (6) discuss the fundamental questions that the estimand framework raises for planning, conducting and analyzing non-inferiority trials. A key conclusion of Lynggaard *et al.* was that the "choice of non-inferiority margin must reflect the chosen estimand". This paper investigates in more detail how the estimand impacts the choice of the non-inferiority margin.

To our knowledge, there are no publications that discuss in detail how to derive the non-inferiority margin considering the estimand framework. Lynggaard *et al.* (6) discusses general considerations for estimands in non-inferiority trials including reflections on the impact of estimands on the non-inferiority margin. Morgan *et al.* (7), White *et al.* (8) and Tweed *et al.* (9) are all recent papers that discuss non-inferiority trials from an estimand perspective, but without discussing how the estimand impacts the choice of non-inferiority margin.

Current regulatory guidance (10–12) on derivation of the non-inferiority margin predates ICH E9(R1), and consequently, the impact of the estimand on the non-inferiority margin is not reflected in the guidance. However, in November 2025, the EMA released a draft guideline on non-inferiority and equivalence comparisons in clinical trials (13). In this draft guideline, a section is dedicated to the selection of the non-inferiority margin, and it is emphasized that the estimand in the historical trials is to be considered. Furthermore, it states that clinical and statistical input is needed to mitigate the likely violation of the constancy assumption, for example, with respect to slightly different populations, changes in standard-of-care and different intercurrent event frequencies that may occur over time. Two methods for deriving the non-inferiority margin are discussed – the fixed margin approach and the synthesis approach. Both methods can be used to demonstrate absolute efficacy (indirect superiority of test treatment versus placebo), but only the fixed margin approach can be used to demonstrate relative efficacy (reference treatment can safely be replaced by test treatment).

Like the EMA draft guideline, the current FDA guidance on non-inferiority trials (10) describes in detail the two different approaches for deriving the non-inferiority margin. However, the guidance is from 2016 and, consequently, does not consider how estimands impact the derivation. For the fixed margin method, which in the experience of the authors, is the more common method, the derivation consists of two steps: a statistical step followed by a clinical step. The FDA guidance states that the effect of the reference treatment versus placebo must be estimated by doing a meta-analysis of historical trials that compare the reference treatment to placebo (statistical step). This effect is named $M_1$. The estimated size of $M_1$ depends on the specific estimand and, in particular, to the strategies for handling intercurrent events in historical trials (6). The non-inferiority margin is determined as the largest loss of effect of the test treatment versus the reference treatment that is clinically acceptable (clinical step). This is denoted as $M_2$. The margin $M_2$ cannot be larger than $M_1$, and usually, it will be smaller than $M_1$. Demonstrating non-inferiority with respect to $M_1$ would, in general, not demonstrate that the test treatment retains enough of the effect of the reference treatment.

This paper investigates to what degree $M_1$ depends on the intercurrent event strategies both through simulation (Section 2) and by use of examples from weight management (Sections 3 and 4). Section 3 focuses on the setting when historical trials of the reference treatment use the estimand framework. Section 4 investigates the scenario when the historical trials do not employ the estimand framework. We discuss the findings of this paper in Section 5 and conclude with recommendations on the choice of non-inferiority margins in light of the estimand framework (Section 5).

# 2 The estimand specificity of the non-inferiority margin

The following section extends the discussion from Lynggaard *et al* (6) on how the specification of non-inferiority margin is always with respect to a particular estimand and quantifies the effect of the choice of the estimand on the non-inferiority margin.

Non-inferiority is always with regard to a specific estimand. For instance, in a trial for a weight management treatment, if the intercurrent event "use of other anti-obesity intervention" is handled using a treatment policy strategy, non-inferiority would be shown for treatment conditions that include the effect of the usage of other anti-obesity intervention. In other words, it would be shown that the test treatment, regardless of usage of other anti-obesity intervention, is non-inferior to reference treatment, regardless of usage of other anti-obesity intervention. Conversely, if a hypothetical strategy is employed, the treatment conditions for which non-inferiority would be shown might be in the setting "had other anti-obesity interventions not been available".

As outlined in the introduction, the non-inferiority margin represents the largest clinically acceptable loss of effect of the test treatment compared to the reference treatment. The size of the loss that is clinically acceptable depends on various factors, including the targeted population, the treatment conditions, the endpoint, and the population-level summary. Using the setting of a weight management treatment as an example, the clinically acceptable loss of effect might differ between obese patients with or without diabetes, between whether or not the treatment conditions include the effect of other anti-obesity intervention, and between relative weight reduction measured at week 52 or 104, or between whether a mean difference or a mean ratio is used to compare treatments. In other words, the largest clinically acceptable loss depends on the estimand of interest. A key factor in defining the estimand, and thus influencing the margin, is the handling strategy of intercurrent events. Consequently, the non-inferiority margin must be defined *a priori* for each estimand and the selection of the historical trials to determine the margin $M_1$, which will later contribute to determining the non-inferiority margin, should ideally take the estimand into account.

Even if various clinical trials formally target the same estimand including applying the same intercurrent event strategies, measured and unmeasured factors might still change over time, leading to differences in the underlying estimand and affecting the relevance of a trial for determining the margin $M_1$. For example, in the setting of an estimand where intercurrent events are handled with a treatment policy strategy, shifts in the frequency of an intercurrent event or different meaning of the intercurrent event (e.g., changes to rescue medication or standard of care) over time can alter the treatment conditions. New intercurrent events may also emerge, potentially becoming part of the treatment condition, and as such altering the definition of the treatment conditions. It is important to emphasize that the magnitude of these influences on the resulting non-inferiority margin is case-specific. In some scenarios, the impact may be negligible, while in others, it could be substantial and should not be overlooked. Therefore, the historical trials which inform the margin $M_1$ and the non-inferiority margin $M_2$, would ideally address the same estimand and define intercurrent events in the same way as the planned non-inferiority trial.

To illustrate how different intercurrent event strategies and variations in the intercurrent event frequency can affect the estimand, and consequently the estimate, we will generate patient journeys of a whole population under a reference treatment and under a placebo. We consider a single intercurrent event and a single occurrence to be part of the patient journey.

The endpoint of interest is the relative percentage weight reduction to baseline. This is measured at 11 post-baseline visits at weeks 4, 8, …, 20, 28, …, 68, i.e., every four weeks until week 20 and every eight weeks thereafter until week 68. We denote the visits by $v = 0,1, …, 11$ with $v = 0$ indicating the baseline visit.

Let $\mathbf{Y}_i = (Y_{i,1}, …, Y_{i,11})$ be the relative percentage weight reduction to baseline in the absence of the intercurrent event for patient $i$. We model $\mathbf{Y}_i$ as multivariate normally distributed, that is, $\mathbf{Y}_i \sim MN(\boldsymbol{\mu}, \boldsymbol{\Sigma})$. The choice of the mean vector in the reference treatment group and the placebo group, respectively, as well as the choice of the covariance matrix in each group are motivated by STEP 1. The detailed parameter choices are listed in Section S1 in the supporting information. For each patient $i$ at each visit $v$, the probability of the intercurrent event is modelled as Bernoulli distributed with probability $p_{i,v}$ with is defined through $\text{logit}(p_{i,v}) = \beta_{0,v} + \beta_1 Y_{i,v}$. We set $\beta_1 = \log(1.1)$ and $\beta_{0,1} = \cdots = \beta_{0,11} = \beta_0$ identical for both trial arms and vary $\beta_0 = \text{logit}(0), \ldots, \text{logit}(0.1)$. The value $\text{expit}(\beta_0)$ can be interpreted as the intercurrent event probability after each visit for a patient with $Y_{i,v} = 0$. The intercurrent event is irreversible, that is, once a patient experiences it, they will no longer be considered free of the event at any subsequent visit. For the post-intercurrent-event behavior, we assume that patients' mean trajectory follows the mean trajectory of placebo group patients, regardless of their initial treatment. The data was simulated using the *simulate_data()* function from the R package {rbmi} (14). We simulate the patient journeys under the reference treatment and placebo for 150000 patients, respectively, and calculate the within-group means and the treatment effect.

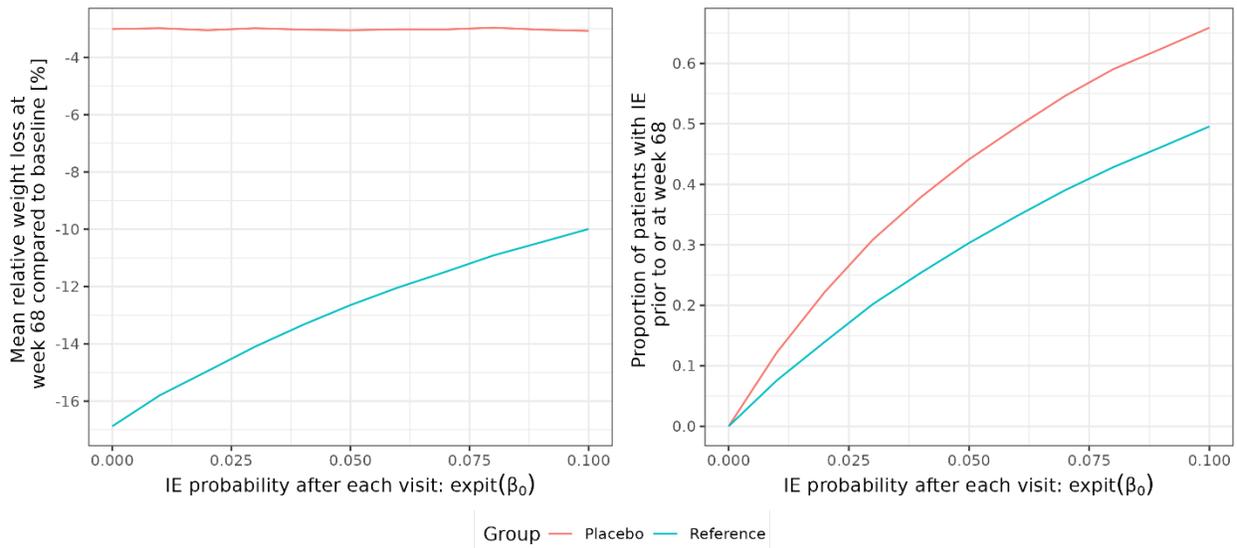

*Figure 1 (left) Group mean at week 68 for the placebo group (red) and the reference group (blue) against the parameter $expit(\beta_0)$. (right) Proportion of patients who have experienced the intercurrent event (IE) prior to or at week 68 against the parameter $expit(\beta_0)$.*

The left panel in Figure 1 shows the group mean at week 68 for the placebo group (red) and the reference group (blue) against the parameter $\text{expit}(\beta_0)$. The right panel in Figure 1 shows the proportion of patients who have experienced the intercurrent event prior to or at week 68 against the parameter $\text{expit}(\beta_0)$. The figures indicate that the group mean in the placebo group remains constant even as the probability of an intercurrent event prior to or at week 68 increases. This is in alignment with the assumed post-intercurrent-event behavior, that is, patients in the placebo arm who experience the intercurrent events behave like patients in the placebo arm who do not experience the intercurrent event. The reference arm shows that an increasing intercurrent event probability decreases the mean relative weight loss at week 68 compared to baseline as more patients behave like placebo patients.

The scenario $\text{expit}(\beta_0) = 0$ is the setting in which no patients have any intercurrent events. In other words, this scenario is the hypothetical estimand 'would patients not experience the intercurrent event' and also the scenario where the hypothetical estimand and the treatment policy estimand are identical. The scenarios for $expit(x) > 0$ correspond to various intercurrent event rates for a treatment policy estimand. Thus, Figure 1 highlights that the treatment effect on the mean relative weight loss at week 68 compared to baseline differs not only between the hypothetical estimand and a treatment policy estimand, but also that for a treatment policy estimand, the estimated value of the population-level summary changes depending on the intercurrent event frequency.

These results emphasize that the (historical) treatment effect of the reference treatment versus placebo, and consequently the margin $M_1$, is specific to an estimand. Moreover, even if different studies assess the treatment policy effect of the experimental treatment versus placebo, they might not necessarily target the same value if the frequency of the intercurrent event affects the treatment effect.

In many historical trials of reference treatment versus placebo, the targeted estimand is not explicitly defined and it may not be possible to determine the exact underlying estimand from reported results and the statistical methods applied. Furthermore, information on the intercurrent events distribution is often not available even in cases where the estimand has been defined, so the validity of pooling historical trials for determining the non-inferiority margin may be unknown.

Even in cases where the estimand is known or can be reasonably inferred from the historical trial(s), there may still be challenges related to assay sensitivity. Assay sensitivity is a property of a clinical trial defined as the ability to distinguish an effective treatment from a less effective or ineffective treatment (15) and justification involves discussion of three aspects: historical evidence of efficacy of the reference treatment, constancy assumption and high quality of trial conduct. Assay sensitivity may be violated if the targeted estimand in the historical trials differs from the estimand that is considered relevant in the non-inferiority trial as a change in any of the estimand attributes would call into question the assumption of historical evidence of efficacy of the reference treatment and the constancy assumption.

# 3 Choice of the non-inferiority margin when historical trials use the estimand framework

In this section, we illustrate practical considerations for determining the non-inferiority margin $M_1$ in a scenario where the clinical efficacy studies for the reference treatment include estimand specifications, but where the estimand of interest in the non-inferiority trial is not estimated in the historical trials of the reference treatment, and there is no access to individual participant data.

We continue to focus on the setting of clinical trials testing a weight management treatment in obese patients. The aim is to determine $M_1$, and these choices are not recommendations for the estimand or the non-inferiority margin in obesity non-inferiority trials.

In Section 3.1 we introduce a fictitious non-inferiority trial where the objective is to demonstrate absolute efficacy of a new weight management treatment and where the fixed-margin approach is used to set the basis for the non-inferiority margin. In Section 3.2, the historical evidence for the reference treatment from the non-inferiority trial will be summarized. In Section 3.3, considerations for determining the margin $M_1$.

## 3.1 Introduction of new non-inferiority trial

The objective of this new non-inferiority trial is to demonstrate absolute efficacy of a new weight management treatment. The reference treatment is once-weekly subcutaneous semaglutide 2.4 mg (in the remainder of the paper, we shall refer to this as semaglutide), which is approved for weight management. The new treatment is expected to have similar efficacy to the reference treatment; consequently, a non-inferiority trial with semaglutide as reference treatment is considered the appropriate choice.

Non-inferiority trials within weight management have not been relevant until now, and so there is no regulatory guidance on a required non-inferiority margin. Thus, in this example, there is a need to derive the margin to be used in the non-inferiority trial.

Treatment with semaglutide is initiated in combination with lifestyle interventions (diet and physical activity counselling) at a dose of 0.25 mg once weekly for the first four weeks and the dose is increased every 4 weeks to the maintenance dose of 2.4 mg at week 16. According to the FDA draft guideline on Obesity and Overweight: Developing Drugs and Biological Products for Weight Reduction (16), the duration of the maintenance period should be at least 52 weeks.

For the primary objective of the non-inferiority trial, the following estimand (clinical question of interest) is proposed:

*What is the difference in mean relative change in body weight (%) from baseline to week 68 between test treatment and once-weekly subcutaneous semaglutide (2.4 mg) both as adjunct to diet and exercise interventions regardless of discontinuation of treatment and as though other anti-obesity interventions were not available in adult patients with a body mass index (BMI) ⩾30 kg/m² (or ⩾27 kg/m² if patients had at least weight-related co-existing condition)?*

The individual estimand attributes are shown in Table 1. In the planned non-inferiority trial, two intercurrent events are anticipated. The intercurrent event "treatment discontinuation" will be addressed with the treatment policy strategy, and the intercurrent event "use of other anti-obesity intervention" by a hypothetical strategy focusing on the setting as if other anti-obesity interventions were not available. That is, we are estimating the effect of prescribing treatment had anti-obesity interventions not been available, applying this hypothetical strategy attempts to assess the causal effect of the medicine itself (17) rather than the effect of the medicine in combination with varied use of rescue medication.

*Table 1 Estimand of interest in non-inferiority trial.*

| Attribute | Description |
|---|---|
| Target population | Adult patients with a BMI ≥30 kg/m² (or ≥27 kg/m² if patients had at least one weight-related co-existing condition) |
| Endpoint | Relative change in body weight (%) from baseline to week 68 |
| Treatment conditions | New treatment versus once-weekly subcutaneous semaglutide (2.4 mg) both as adjunct to diet and exercise interventions regardless of discontinuation of treatment and as though other anti-obesity interventions were not available |
| Population-level summary | Mean relative change in body weight (%) from baseline to week 68. The between-group comparison is based on the mean difference. |
| Intercurrent events and handling strategy | • Treatment discontinuation: Treatment policy (addressed in the treatment conditions attribute)<br>• Other anti-obesity intervention: Hypothetical - as though other anti-obesity interventions were not available (addressed in the treatment conditions attribute) |

## 3.2 Overview of historical evidence for reference treatment

A key part of determining the non-inferiority margin is understanding the treatment effect of semaglutide, i.e., the reference treatment. The safety and efficacy of semaglutide for weight management was assessed in the STEP clinical trial program. This program includes to date eleven phase 3 clinical efficacy studies, one of which was in teenagers (STEP TEENS), and a phase 2 dose-ranging trial (in the remaining referred to as STEP 0). The information presented in this section is solely based on the published results of the STEP trials (18–29), and without access to individual participant data.

The estimand framework was implemented in all trial protocols to precisely define the clinical questions of interest.

All STEP trials compared treatment conditions based on semaglutide to treatment conditions based on placebo, the precise definitions of the treatment conditions attribute in each trial are provided in supporting information Table S2. The primary bodyweight endpoint recommended by FDA (16) is percentage change in bodyweight from baseline to a specified timepoint, which is at least one year after treatment assignment. This endpoint was in the STEP trials typically assessed between 52 and 68 weeks, and the basis for comparing treatment conditions was the difference in mean relative change.

The target population differed between trials, see Table S2. Most trials included adults with overweight (BMI ⩾27 with at least one weight-related comorbidity) or obesity (BMI⩾30), but STEP 9 included adults with obesity and a clinical and radiologic diagnosis of moderate knee osteoarthritis with at least moderate pain, and STEP TEENS included adolescents (12 to <18 years of age) with obesity (a BMI in the 95th percentile or higher) or with overweight (a BMI in the 85th percentile or higher) and at least one weight-related coexisting condition.

The following intercurrent events were identified in the STEP studies in relation to the bodyweight objective:

- Discontinuation of investigational medicinal product (IMP).
- Use of other anti-obesity interventions, such as anti-obesity medication or bariatric surgery.

The clinical trials from the STEP program all reported two different treatment effects (estimands): an estimand where both intercurrent events are handled by the treatment policy strategy and an estimand where both intercurrent events are handled by a hypothetical strategy. A generalized wording of the two estimands defined in the STEP trials is presented in Table 2. Further details from the respective publications can be found in Section S2.1 in the supporting information.

Table 2 Overview of estimands in the STEP trials.

| Attribute | Primary estimand | Supplementary estimand |
|---|---|---|
| Target population | Adult patients with a BMI ⩾30 kg/m$^2$ (or ⩾27 kg/m$^2$ if patients had ⩾1 weight-related co-existing condition). | |
| Endpoint | Relative change from baseline to week 68 in body weight (%). | |
| Treatment conditions | Once-weekly subcutaneous semaglutide (2.4 mg) versus placebo both as adjunct to diet and exercise interventions regardless of discontinuation of treatment and including the potential effect of other anti-obesity interventions. | Once-weekly subcutaneous semaglutide (2.4 mg) versus placebo both as adjunct to diet and exercise interventions as though treatment was not discontinued and as though other anti-obesity medication was not available. |
| Population-level summary | The between-group comparison is based on the difference of the mean relative change in body weight (%) from baseline to week 68. | |
| Intercurrent events and handling strategy | Treatment discontinuation: Treatment policy (addressed in the treatment conditions attribute) Other anti-obesity intervention: Treatment policy (addressed in the treatment conditions attribute) | Treatment discontinuation: Hypothetical – as though treatment was not discontinued (addressed in the treatment conditions attribute) Other anti-obesity intervention: Hypothetical - as though other anti-obesity intervention was not available |

| | | (addressed in the treatment conditions attribute) |
|---|---|---|

The estimated treatment effects for the two estimands defined in Table 2 for STEP 1 to STEP 10 trials are shown in Figure 2 as well as in Table S2. STEP 0 and STEP TEENS are excluded from the figure, since STEP 0 included different doses of semaglutide with a once-daily administration, and STEP TEENS used a different method of assessment of weight change to mitigate the effect of growing.

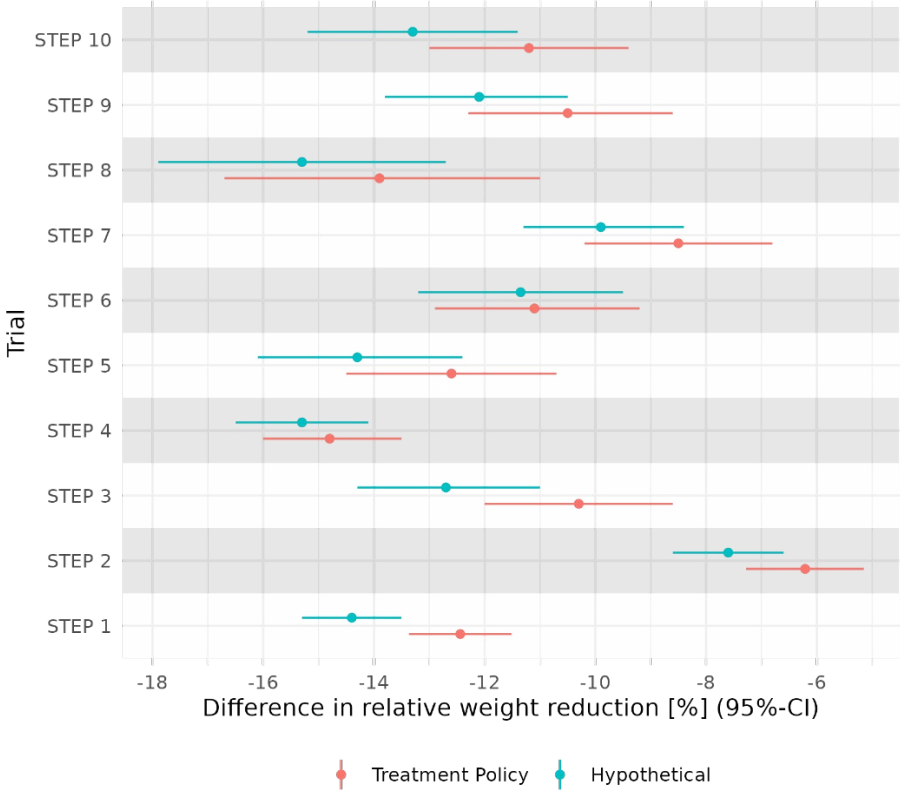

*Figure 2 Estimated treatment effects with 95% confidence intervals for the two estimands defined in the STEP trials.*

Figure 2 suggests some degree of heterogeneity in the treatment effect estimated across the trials despite the similarity of estimands across trials. Moreover, Figure 2 shows that the estimated treatment effect for the treatment policy estimand is consistently smaller than the estimated treatment effect for the hypothetical estimand. Substantial heterogeneity questions the validity of pooling trials. Two potential sources of variability in weight management are recognized: type 2 diabetes and gender.

Being overweight is associated with type 2 diabetes, and it is well-known that the effect of weight reduction treatments is less efficacious in patients with type 2 diabetes (16) compared with individuals without diabetes. The largest change in bodyweight is seen in normo-glycemic individuals and the smallest in patients with type 2 diabetes. In addition, individuals with pre-diabetes have a smaller weight reduction compared to normo-glycemic individuals (30)]. Some STEP trials excluded patients with type 2 diabetes at baseline, whereas other STEP trials included them; thus, changing the target population. In addition, the effect of semaglutide in women is generally larger than in men (29).Table 3 summarizes the proportion of participants with type 2 diabetes or prediabetes and the gender distribution by each of the STEP trials.

*Table 3 Proportion of participants with type 2 diabetes or prediabetes and gender distribution by trial.*

| Trial | Inclusion of participants with type 2 diabetes or with prediabetes at baseline? | Gender distribution % females |
|---|---|---|
| STEP 0 | No | 65 |
| STEP 1 | No | 74 |
| STEP 2 | 100% with type 2 diabetes | 51 |
| STEP 3 | No | 88 |
| STEP 4 | No | 79 |
| STEP 5 | None with type 2 diabetes, 46% with prediabetes | 78 |
| STEP 6 | 25% with type 2 diabetes, 22% with prediabetes | 37 |
| STEP 7 | 26% with type 2 diabetes, 29% with prediabetes | 45 |
| STEP 8 | No | 78 |
| STEP 9 | None with type 2 diabetes; prediabetes not reported | 82 |
| STEP 10 | 100% with prediabetes | 71 |
| STEP TEENS | 4% with type 2 diabetes | 62 |

## 3.3 Considerations for determining the margin $M_1$

In this section, we illustrate practical considerations for determining the margin $M_1$ based on the historical evidence presented in Section 3.2. Each of the historical studies estimated two estimands: a treatment policy estimand and a hypothetical estimand, confer Table 2. However, the planned non-inferiority trial addresses one intercurrent event with a treatment policy strategy and another with a hypothetical strategy, respectively, confer Table 1. Thus, the key challenge is that the estimand of interest in the non-inferiority trial is not estimated in the historical trials of the reference treatment, and no individual patient data is available. Next, we perform a meta-analysis for each of the estimands to calculate the margin $M_1$ for each of the estimand and to characterize the differences in the margin between the estimands. Thereto, the relevant STEP trials must be selected. Note, the intention of this section is not to discuss how to do a meta-analysis, but rather to illustrate how the estimand impacts the size of the estimated treatment effect from the meta-analysis. Consequently, all aspects of good reporting practice as recommended in the PRISMA statement (31) are not addressed in this paper.

Twelve STEP trials were available to us. The following selection criteria were applied:

- Population: Adults with overweight (BMI>= 27 with at least one weight-related comorbidity) or obesity (BMI>=30) – with or without type 2 diabetes at baseline
- Dosing regimen: once-weekly s.c. semaglutide 2.4 mg
- Endpoint: Relative change in body weight (%) from baseline to a timepoint within 52-68 weeks after baseline.

Due to the difference in treatment effects between patients with and without type 2 diabetes as discussed in Section 3.2, in practice one might want to conduct two separate meta-analyses. However, for the purpose of illustration, the focus in this section will be on setting up one non-inferiority trial that includes the broader population and, hence, only conducting a single meta-analysis to determine the treatment effect of the reference treatment, that is $M_1$.

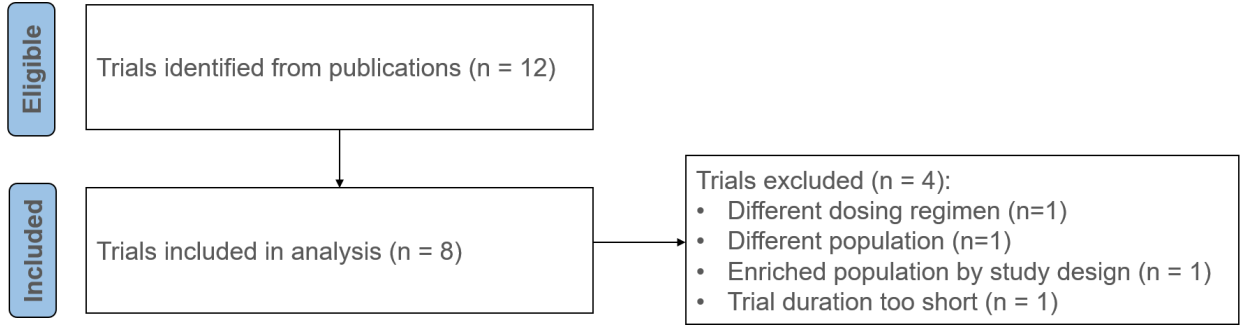

Figure 3 Selection of STEP trials for the meta-analysis.

Of the twelve STEP trials, eight could be included in the meta-analysis, (Figure 3): Four trials were excluded: STEP 0 (different doses and once-daily administration), STEP TEENS (different population and because assessment of weight change is related to body mass index), STEP 4 (treatment effect in tolerators of semaglutide) and STEP 7 (44-week exposure).

Some of the STEP trials include liraglutide 3.0 mg or different doses of semaglutide. The non-inferiority trial will use once-weekly s.c. semaglutide 2.4 mg as reference treatment. Consequently, the meta-analysis included information on that dose only. The number of participants assigned to that dose and to placebo in the STEP trials can be found in Section S1.

The non-inferiority margin $M_1$ was determined through a Bayesian meta-analysis using the R package {RBesT}. In detail, let $y_i, i = 1, \ldots, H$, be the treatment effect estimates from $H$ historical trials and let $s_i^2$ be the corresponding estimated variance of the treatment effect estimates. A standard normal-normal hierarchical model is assumed, i.e., $y_i|\theta_i \sim N(\theta_i, s_i^2)$ with $\theta_i \sim N(\mu, \tau^2)$. The parameter $\mu$ is the overall treatment effect. The R code for the meta-analysis is provided in Section S3. The results from the meta-analysis are shown in Table 4.

*Table 4 Estimated mean treatment effects and 95% credible intervals from the meta-analysis of the selected STEP trials.*

| Handling strategy for discontinuation and other anti-obesity intervention | Pooled % mean treatment effect and 95% credible interval |
|---|---|
| Treatment policy | -10.9 (-13, -8.85) |
| Hypothetical | -12.6 (-14.8, -10.3) |

The estimated mean difference (95% credible interval) of once-weekly subcutaneous semaglutide (2.4 mg) versus placebo, both as adjunct to diet and exercise interventions regardless of discontinuation of treatment and including the potential effect of other anti-obesity interventions, is -10.9% (-13%, -8.85%). Thus, the margin $M_1$ in terms of a mean difference between the new treatment and semaglutide, both as adjunct to diet and exercise interventions regardless of discontinuation of treatment and including the potential effect of other anti-obesity interventions, to be ruled out is 8.85%.

The estimated mean difference (95% credibility interval) of once-weekly subcutaneous semaglutide (2.4 mg) versus placebo, both as adjunct to diet and exercise interventions as though treatment was not discontinued and as though other anti-obesity medication was not available, is -12.6% (-14.8%, -10.3%). Thus, the margin $M_1$ in terms of a mean difference between the new treatment and semaglutide, both as adjunct to diet and exercise interventions as though treatment was not discontinued and as though other anti-obesity medication was not available, to be ruled out is 10.3%.

Figure 2 suggests that STEP 2 may be an outlier, so a sensitivity analysis was made excluding this trial. The results are presented in Supporting Information, Section S.2.2.2. The estimated treatment effects were slightly larger compared to the main results.

As outlined in Table 1, in the new non-inferiority trial, the reference treatment condition is defined as semaglutide as adjunct to diet and exercise interventions regardless of discontinuation of treatment and as though other anti-obesity interventions were not available. The effect of this reference treatment condition was not studied in the STEP program, i.e., the STEP program handled the intercurrent events (discontinuation of treatment, use of other anti-obesity interventions) with different strategies than the new non-inferiority trial. Thus, the effect of reference treatment would most likely lie between the two $M_1$ margins, that is, between 10.3 and 8.85. A practical choice in this situation could be to take forward the smaller value $M_1$ of 8.85 to discuss the non-inferiority margin with clinicians. Note, this should not be regarded as a recommendation for always using the results from estimating a treatment effect based on the treatment policy strategy as the choice will be context dependent.

# 4 Choice of the non-inferiority margin when historical trials do not employ the estimand framework

## 4.1 General considerations

Prior to the release of the ICH E9(R1) addendum, there was no requirement to specify the estimand, as this was not standard practice at the time. Therefore, when considering relevant historical clinical trials against the targeted estimands when designing a new non-inferiority trial, the estimand from these historical trials are not always (fully) specified in trial documents (e.g., clinical trial protocol, statistical analysis plan, etc.) or the trial publications, and often the available information is not sufficient to reconstruct the targeted estimand. Therefore, determining an appropriate non-inferiority margin presents a significant challenge, since there is no standardized guidance on how the non-inferiority margin should be established within the estimand framework. In this section, various sources of information that can potentially help identify the estimand will be discussed.

Ideally, where it is possible to access individual participant data (IPD) of relevant trials (either internally or by use of open-access data request sources (e.g., clinicalstudydatarequest.com, data.projectdatasphere.org), retrospective estimation of the estimand of interest should be undertaken. This approach is only feasible if data on relevant intercurrent events were collected. In practice, however, obtaining and analyzing IPD via open-access requests can be time-consuming and may be practically infeasible.

When complete information to determine the estimand is unavailable from published sources and all the relevant IPD cannot be accessed, assumptions must be made about how any intercurrent events were addressed in historical studies. Wherever possible, informed judgments should be made about the likely estimand for each trial to be included in the meta-analysis to derive $M_1$, ensuring any assumptions made are plausible and transparently documented. This process may involve an in-depth review of the clinical trial protocol, statistical analysis plan or clinical trial reports available in clinical trials registers (e.g., clinicaltrials.gov), the European Public Assessment Report (EPAR) or using the CONSORT flow diagrams available within publications. For example, the CONSORT disposition flow charts can be screened for information on potential intercurrent events. The additional information can help provide context for the trial and clarify how the trial was conducted, what data were collected, and how analyses were performed.

In particular, it is important to consider what the standard statistical approach to analyze the data was at the time of the planning and conduct of the historical clinical trials. In addition, it is often helpful to understand the clinical setting at the time and which (disease-specific) regulatory guidelines were effective.

To retrospectively identify the targeted estimand, it is essential to understand how post-intercurrent-event data are handled in the statistical analysis. If post-intercurrent-event data are considered relevant to the treatment effect of interest and are included in the analysis, the targeted estimand is generally closely aligned with a treatment policy estimand, and the emphasis shifts to the appropriateness of the missing data handling. If post-intercurrent-event data are treated as missing data and subsequently modelled as if the intercurrent event did not occur, the targeted estimand is generally aligned with a hypothetical estimand, and the key question is which hypothetical scenario the modelling corresponds to. Post-intercurrent-event data are excluded, and the endpoint is modified to incorporate the occurrence of the intercurrent event into the definition of the endpoint, the targeted estimand corresponds to a composite estimand. An analysis restricted to pre-intercurrent-event data, without modelling or imputing post-intercurrent-event data, targets an estimand generally closely aligned with a while-alive estimand. If the analysis restricts estimation to participants defined by a post-randomization characteristic related to the intercurrent event (e.g., those who would not experience the event under any treatment) and focuses on the treatment effect within that subpopulation, then the targeted estimand may be closely related to a principal stratum estimand. Finally, some analyses may, even under reasonable assumptions, not align with a causal estimand (e.g., analyses based on the per-protocol set) and some analyses might align with multiple estimands (e.g., when no post-intercurrent-data has been collected).

An overview of the process is provided in Figure 4.

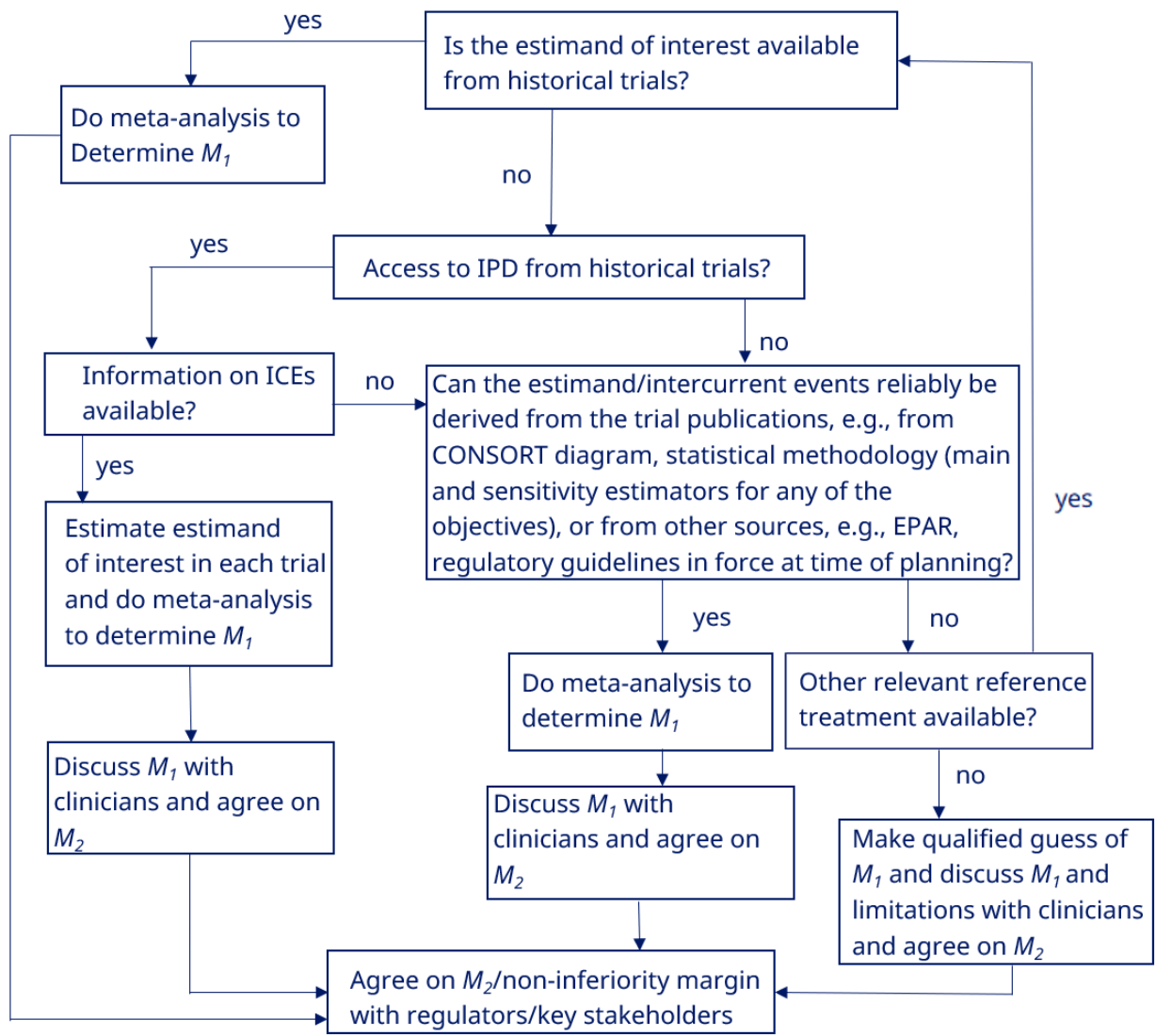

*Figure 4 Overview of process for deriving the estimand in historical trials. EPAR: European Public Assessment Report.*

To illustrate how this process could be applied when there is incomplete information on the estimand from historical studies, we first introduce our second motivating example. We then outline the logic of retrospectively describing the estimand to enable the derivation of $M_1$, followed by key practical considerations and highlight any learnings.

## 4.2  Motivating example: A new weight management treatment versus liraglutide

Consider a new weight loss trial where the objective is to demonstrate non-inferiority of a new treatment compared with daily subcutaneous liraglutide 3.0 mg. The new treatment is expected to have similar efficacy to liraglutide. All attributes of the estimand for this trial are similar to the estimand described in Table 1, except for the treatment conditions as the assigned treatments are test treatment versus daily subcutaneous liraglutide (3.0 mg) both as adjunct to diet and exercise interventions. Therefore, the clinical question of interest is:

*What is the difference in mean relative change in body weight (%) from baseline to week 68 between test treatment and once-daily subcutaneous liraglutide (3.0 mg) both as adjunct to diet and exercise interventions regardless of discontinuation of treatment and as though other anti-obesity interventions were not available in adult patients with a body mass index (BMI) ⩾30 kg/m² (or ⩾27 kg/m² if patients had at least weight-related co-existing condition)?*

To determine $M_1$ for our new trial, relevant historical trials that assessed the efficacy and safety of liraglutide (3.0 mg) were identified. For simplicity, we focus on the SCALE clinical development program which included four phase 3 efficacy trials comparing liraglutide (3.0 mg) with placebo (32–35). This program focused on weight management in patients who are overweight (BMI $\geq 27\ kg/m^2$) or obese (BMI $\geq 30\ kg/m^2$) with additional comorbidities (hypertension, type 2 diabetes, dyslipidemia and obstructive sleep apnea). The main objective was to demonstrate that liraglutide (3.0 mg) would result in clinically meaningful weight loss in comparison to placebo. Of these four trials, only the SCALE Insulin trial (32) had fully implemented the estimand framework. The other three trials were published prior to the finalization of ICH E9(R1) (1).

## 4.3   Overview of historical evidence for reference treatment

The first thing to consider is the context in which the historical trials were planned and conducted.

The SCALE articles reporting the results from the four SCALE trials we have identified were published in the period from 2013-2020. During this time-period, ICH E9(R1) was finalized (November 2019), and the FDA guidance on clinical trials in weight management (36) specified that an ANCOVA with last observation carried forward (LOCF) on treatment approach should be followed for the primary analysis. The guidance specified that trial participants who were withdrawn from the trial should be called in for the final assessments to be made at the landmark visit, but not a full follow-up. According to the guidance, the full analysis set (FAS) was to be defined as all randomized participants who received at least one dose of trial drug and had at least one post-baseline assessment (referred to as modified ITT population). This definition was used in SCALE Obesity and pre-diabetes and SCALE Maintenance trials. The SCALE Sleep Apnea (33)included all randomized participants. It is unclear what the definition was in the SCALE Insulin trial, but the statistical section and the reported results in the publication (32) indicate that all randomized participants were included in the estimation of the two estimands.

At the time when the trials without explicit estimands defined were conducted, treatment discontinuation implied withdrawal from the trial and the specific reason for treatment discontinuation did not impact that decision. Also, an analysis said to be based on the full analysis set (FAS) most often meant that all randomized participants were included in the analysis, but it did not mean that all randomized participants were followed up until the planned end of trial. In addition, at the time when the SCALE trials were conducted, there was a limited number of other anti-obesity interventions (medications and surgery) available, and these were not commonly used. Consequently, use of other anti-obesity intervention is not considered as an intercurrent event in the SCALE trials, but it has been identified in the new non-inferiority trial, implying different treatment conditions This means that assay sensitivity is likely violated through violation of the constancy assumption. This can to some extent be accounted for when discussing and agreeing on the non-inferiority margin, cf. Section 4.3.5. To help determine $M_1$, we scrutinized the published articles including published supplementary material to extract any useful information on the estimand attributes and on the estimators used. The results from the primary analysis, sensitivity analyses or secondary analyses in relation to the percentage change in body weight are presented in Supporting Information, Table S10. We examined both main analyses and any sensitivity analyses to determine if there were common approaches across trials. From the reported results of the SCALE trials, we compared the number of randomized patients (or in FAS) and the number of patients included in the analysis (where available).

The purpose of this process is to make informed and transparent assumptions about the (likely) targeted estimand(s) in each trial, enabling an informed selection of the most relevant results for a meta-analysis to determine $M_1$.

Details on the derived estimand information are provided in Supporting Information, S3 and summarized below. The key intercurrent event to be considered is treatment discontinuation for all trials, the target population, treatment conditions, endpoint and population-level summary are provided before considering this intercurrent event. Depending on the strategy/strategies used, the first three attributes would be modified.

### 4.3.1 SCALE Maintenance Trial

The main results of the Scale Maintenance trial are published in (35). Information on some of the estimand attributes is (partly) available from the main paper:
- **Target population:** adult patients with a body mass index (BMI) ≥30 kg/m2 (or ≥27 kg/m2 if patients had dyslipidemia or hypertension) and without type 1 or type 2 diabetes.
- **Treatment conditions:** once-daily subcutaneous liraglutide (3.0 mg) versus placebo both as adjunct to diet and exercise interventions.
- **Endpoint:** Relative (percentage) change from baseline to week 56 in fasting body weight.
- **Population-level summary:** Difference in mean relative changes in fasting body weight from baseline to week 56.

The primary analysis used an ANCOVA model with a last observation carried forward (LOCF) approach to deal with missing data in alignment with the FDA guidance. It is unclear whether the data retrieved after treatment discontinuation were used in the analysis and whether only truly missing data were imputed by LOCF, but since the FDA guidance states, "last observation carried forward on treatment", and since one of the specified sensitivity analyses did include retrieved data, we assume that the primary analysis did not. The primary approach does not clearly and immediately align with any specific handling strategy, so we further examined the reported sensitivity analyses.

The trial reported several sensitivity analyses, including a mixed model for repeated measures (MMRM) analysis. Supplementary Table 2 within the article's Supplementary Information states in the footnotes of the results: The repeated measures analysis (linear mixed-effect model) was done using the full analysis set with no imputation. Together with the number of participants included in the analysis and the numbers reported in the CONSORT diagram, it suggests an MMRM on the sets of participants who completed the trial. This does not align with any specific strategy.

Another sensitivity analysis included was an ANCOVA using the retrieved data and LOCF for truly missing data. This would align with a treatment policy strategy.

### 4.3.2 SCALE Obesity and Prediabetes Trial

The results from the SCALE Obesity and prediabetes trial are published in Pi-Sunyer *et al* (34). Information on some of the estimand attributes is (partly) available from the main paper:
- **Target population:** adult patients with a body mass index (BMI) ≥30 kg/m2 (or ≥27 kg/m2 if patients had dyslipidemia or hypertension) without type 1 and type 2 diabetes.
- **Treatment conditions:** once-daily subcutaneous liraglutide (3.0 mg) versus placebo both as adjunct to diet and exercise interventions.
- **Endpoint:** Relative (percentage) change from baseline to week 56 in fasting body weight.
- **Population-level summary:** difference in mean relative changes from baseline to week 56.

The primary analysis again used ANCOVA with last observation on treatment carried forward imputation and several sensitivity analyses were reported (see Table S8 in the Supporting Information). One sensitivity analysis used MMRM, which could be interpreted as estimating a hypothetical strategy to handle treatment discontinuation depending on whether or not the off-treatment data were used. It is unclear whether this was the case. It should be noted that several participants were excluded from this analysis, most likely due to no trial drug exposure or because no endpoint data were available to calculate a change from baseline. Like for the SCALE Maintenance trial, a sensitivity analysis similar to the primary analysis where all available data were included and truly missing data was imputed by LOCF, was specified. Again, this would reflect a treatment policy strategy.

### 4.3.3 SCALE Sleep Apnea Trial

The results from the SCALE Sleep Apnea trial are published in Blackman *et al* (33). Information on some of the estimand attributes is (partly) available from the main paper:
- **Target population:** adult patients with a body mass index (BMI) ≥30 kg/m2 diagnosed with moderate or severe obstructive sleep apnea, and without type 1 and type 2 diabetes.
- **Treatment conditions:** once-daily subcutaneous liraglutide (3.0 mg) versus placebo both as adjunct to diet and exercise interventions.
- **Endpoint (secondary):** Relative (percentage) change from baseline to week 32 in fasting body weight.
- **Population-level summary:** difference in mean relative changes from baseline to week 32.

The primary objective of the SCALE sleep apnea trial is not related to bodyweight change, but a secondary objective with a secondary endpoint being relative change from baseline in bodyweight. Unfortunately, due to the different primary objectives, the duration of the trial is only 32 weeks, which is too short to see the full effect on bodyweight. Consequently, the trial does not add relevant information for the estimand of interest in the new non-inferiority trial and will therefore not be used in the meta-analysis. Furthermore, the population is (slightly) different from the one in the non-inferiority trial. However, this trial illustrates the importance of considering all objectives and endpoints and not just the primary, since it is likely that the trial still can contribute with information in case the endpoint of interest is defined in the trial.

### 4.3.4 SCALE Insulin Trial

The results from the SCALE Insulin are published in Garvey *et al* (32). Information on some of the estimand attributes is (partly) available from the main paper:
- **Target population:** adult patients with a body mass index (BMI) ≥27 kg/m2 diagnosed with type 2 diabetes and treated with basal insulin and no more than two oral anti-diabetic medications.
- **Treatment conditions:** once-daily subcutaneous liraglutide (3.0 mg) versus placebo both as adjunct to diet and exercise interventions.
- **Endpoint:** Relative (percentage) change from baseline to week 56 in fasting body weight.
- **Population-level summary:** difference in mean relative changes from baseline to week 56.

The SCALE Insulin trial had clearly specified the estimand. The section "Study Design" indicates that a potential intercurrent event was not defined as such: "stop and restart of study drug". However, it can be assumed that it was an intercurrent event that was treated by the treatment policy strategy. Unfortunately, not defining this as an intercurrent event means that no information on the frequency of that is available.

In this trial, treatment discontinuation was handled using a treatment policy strategy for the primary estimand, which was estimated by an ANCOVA with missing data being imputed by a referenced-based imputation. A supplementary estimand was defined, where treatment discontinuation was handled by a hypothetical strategy. The supplementary estimand was estimated by a standard MMRM analysis using on-treatment measurements.

### 4.3.5 Meta-analysis

Based on the information we could extract from the SCALE trial publications and retrospectively determining the estimand targeted, we consider which of those trials we can take forward to help determine $M_1$ via a meta-analysis. We reiterate that the intent of this exercise is for the purpose of illustration with regards to the approach of determining $M_1$ rather than the meta-analysis itself.

The SCALE Sleep apnea trial does not qualify for inclusion in the meta-analysis due to the short trial duration of 32 weeks.

Regarding the remaining three trials, we can use the results where we have inferred that a treatment policy strategy was used to address treatment discontinuation. However, it should be noted that the estimators were different, and the populations were also different with respect to including participants with type 1 and type 2 diabetes, which is known to impact the treatment effect size. We are unable to determine whether a hypothetical strategy was also used to address treatment discontinuation for the SCALE Maintenance and SCALE Obesity and pre-diabetes trials. This is because where an MMRM analysis was used, it is unclear whether the off-treatment data were also included within the analysis.

The results mostly aligned with the treatment policy strategy from the three eligible trials for the meta-analysis are summarized in Table 5. We present the results from the SCALE Insulin trial where a hypothetical strategy was used and we present the results from the SCALE Obesity and pre-diabetes trial where an MMRM analysis was used. The MMRM results for the SCALE Maintenance trial are not included since this analysis is based on those that completed the trial. The number of patients included within the analyses of a trial may influence how much weight a given trial should carry in the overall assessment when deriving $M_1$.

*Table 5 Potential information to use from historical trials.*

| Trial | Results inferred for treatment policy | Results inferred for hypothetical | MMRM results |
|---|---|---|---|
| SCALE Maintenance | -5.4 (-6.8, -3.9) | *Not applicable\** | *Not applicable\** |
| SCALE Obesity and pre-diabetes | –5.2 (–5.6, –4.7) | *Unknown* | –5.8 (–6.3, –5.3) |
| SCALE Insulin | –4.3 (–5.5, –3.2) | –5.1 (–6.3, –3.9) | –5.1 (–6.3, –3.9) |

*Analysis based on completer set

The estimated treatment difference and 95% credible interval from the meta-analysis based on the treatment policy treatment effects in Table 5 were -5.04% and (-6.87, -2.94), respectively. Thus, the $M_1$ to carry forward for discussion with clinicians is -2.94%.

An additional complexity arises when the context of intercurrent events changes over time. For example, anti-obesity medication was not considered an intercurrent event in any of the SCALE studies. In this case, where a relevant intercurrent event for the new trial was not explicitly accounted for historically, discussions on the anticipated frequency of this intercurrent event and the expected impact of the treatment effect should be discussed with clinicians when determining $M_2$. Simulation studies like the one presented in Section 2 could be valuable for exploring the potential impact of these different assumptions. Therefore, when determining $M_1$ using a meta-anlaysis it is essential to consider how broadly or narrowly the population of interest is defined and how relevant the historical data are with respect to the planned trial, particularly when the treatment strategies evolve.

# 5 Discussion

Many of the uncertainties encountered in non-inferiority margin determination arise because historical trials lack explicit descriptions in trial protocols and publications of how treatment effects were defined. The introduction of the ICH E9(R1) estimand framework has enabled a clearer description of intercurrent events and which strategies were used to account for these events. Going forward, transparent and comprehensive reporting of estimands in contemporary trials will help address this problem. Part of the EMA 3-year (2025-2027) methodology working plan (37) is to provide guidance on aligning estimand attributes across different studies in the context of meta-analysis which directly acknowledges the importance of this issue.

Currently, the determination of the $M_1$ margin based on meta-analysis is likely to rely on information from historical trials which were completed and reported prior to the introduction of the ICH E9 (R1) estimand framework (1). The details provided in trial publications may not be sufficient to determine with confidence the intended estimand from these historical trials. Any margin derived from those studies therefore relies on assumptions about how treatment effects were defined unless individual patient data which includes detailed information on intercurrent events is available for re-analysis. The limitations of using historical trials should be explicitly recognized and sensitivity analyses are important to assess the robustness of the results of the meta-analysis to these assumptions.

The identification of historical trials for meta-analysis requires genuine cross-functional collaboration. Clinicians and statisticians should jointly evaluate whether historical trials address the same clinical question and whether their estimands align in meaning with those intended for the new non-inferiority trial. Differences in clinical practice and evolving definitions of intercurrent events may affect the relevance of historical evidence. Assay sensitivity may be compromised if the estimand targeted in the historical trials differs from the estimand deemed relevant for the non-inferiority trial. Changes in any estimand attribute can undermine both the assumption that historical trials provide valid evidence of the reference treatment's efficacy and the constancy assumption. Potential impacts on assay sensitivity should therefore be carefully evaluated when defining the non-inferiority margin. These considerations should be explicitly documented.

The complexities of these evaluations were clearly demonstrated in our second example. Retrospectively determining the estimand and deriving an appropriate choice of $M_1$ from historical trials proved challenging. Although it was possible to extract considerable information by carefully reviewing each trial publication, the relevance of that information to the new trial being designed was not always straightforward. When using published data to derive $M_1$, it is essential to review all available supplementary information, including protocols, analysis plans, and CONSORT flow diagrams, from those historical trials. Guidelines that were effective at the time the historical trials were planned and conducted may provide additional context. For example, the disposition flow diagrams provided useful insights into reasons for trial withdrawal or treatment discontinuation enabling at least identification of potential intercurrent events. One challenge, however, was trying to distinguish whether the reasons listed within the flow diagram would now be classified as intercurrent events under ICH E9(R1) or were based on missing data. While such inferences are necessarily imperfect and the full context may be unavailable, this approach nonetheless supports a reasonable assessment of what occurred and how many patients experienced intercurrent events.

As illustrated through the second example, perfect alignment between historical and planned trials will often not be achievable, and therefore informed assumptions must be made about the estimand in historical trials, especially when information does not exist, e.g., on the intercurrent event. While this is a practical necessity, it is also critical when deriving a defensible choice for $M_1$ and therefore the margin in such challenging scenarios. Critically, any assumptions made must be plausible, transparent, and clearly documented, preferably in the protocol. By laying out the assumptions explicitly, trial teams can make an informed choice for what $M_1$ should be and subsequently, sponsors and regulatory agencies can have a shared basis for discussion. Regulators then can either agree or provide informed feedback on whether the estimated treatment effect of the reference treatment, $M_1$ is acceptable. It is crucial that agreement on the final choice of margin between sponsor and regulator is reached prior to trial initiation.

The first example, based on the STEP studies, illustrated that two different estimands lead to different estimates of $M_1$ (the effect of active compared with placebo). Because $M_2$ (the clinically acceptable margin) will be informed by estimates of $M_1$, differences in $M_1$ would be expected to be reflected in $M_2$. This implies that distinct estimands could reasonably lead to different values for $M_2$ making the use of a single common $M_2$ for both estimands potentially inappropriate and difficult to justify.

The Cochrane Collaboration is a highly respected source of methodology for the conduct of systematic reviews and meta-analyses, and its influence on evidence synthesis is substantial. In this context, it is noteworthy that despite calls for alignment (38,39), the Cochrane handbook for reviews does not acknowledge the ICH E9(R1) framework. As a result, meta-analyses reported by the Cochrane Collaboration may combine treatment effects defined under different implicit strategies, which can complicate their use when deriving non-inferiority margins. Until the principles of ICH E9(R1) are formally incorporated into evidence synthesis methodological standards, researchers will continue to face challenges when relying on meta-analytic results that do not distinguish among estimands or intercurrent event strategies.

Although conservative assumptions are often viewed as a safeguard in non-inferiority trial design, relying automatically on a conservative estimate of $M_1$ is not recommended. An overly conservative $M_1$ may not reflect the true historical effect of the reference treatment and this can lead to an $M_2$ margin that is unreasonably small. In some settings, a very small $M_1$ leads to a non-inferiority margin that makes trials impractical solely because the margin was misaligned with historical evidence. A reasoned, transparent justification of $M_1$ and the subsequent choice of $M_2$ is therefore more appropriate than a default reliance on a conservative value.

Regulatory treatment-disease guidelines also play a substantial role. Some guidelines have been updated to include the estimand framework (e.g., the EMA guideline for treatment or prevention of diabetes mellitus (40)). Reports from current trials that follow these guidelines should contain explicit details on intercurrent events and the strategies used to address them. However, where guidelines recommend a specific non-inferiority margin without identifying the estimand to which the margin applies, inconsistencies can arise. Updating these guidelines to link recommended margins to specific estimands would reduce ambiguity. Agreement with regulators on the appropriate margin should always be obtained before initiating the non-inferiority trial.

This work has two important limitations. Firstly, the examples focus on estimation of $M_1$ and do not explicitly consider how $M_2$ would be defined under the different estimands. As a result, the practical implications for selecting the clinically acceptable margin are not fully explored. Secondly, the discussion is limited to treatment policy and hypothetical strategies for handling intercurrent events; other strategies are not examined.

# 6 Conclusions

Overall, NI margin determination is only as strong as the clarity of the historical evidence on which it relies. When estimands from historical studies are unknown or inconsistently reported, the resulting uncertainty must be acknowledged and addressed through careful documentation and sensitivity analyses. Continued improvement in estimand reporting, supported by greater sharing of individual patient data, will strengthen the evidence base for NI trials and facilitate more robust and reliable margin determination.

We have summarized our recommendations in Table 7. We welcome more discussion on the topic and encourage more publications on case studies.

*Table 6 Recommendations for establishing and reporting non-inferiority margins.*

| |
|---|
| Ensure the non-inferiority margin $M_2$ is explicitly linked to the primary estimand as the historical effect $M_1$ is dependent on the estimands used in previous trials. |
| Conduct cross-functional collaboration to identify historical trials relevant for the meta-analysis. |
| Perform sensitivity analysis to evaluate the robustness of the meta-analysis results addressing the level of certainty and appropriateness of the estimands in historical trials. |
| Document limitations in the determination of the non-inferiority margin, including considerations on the availability and appropriateness of estimands in historical trials. |
| Consider issues of assay sensitivity in relation to the non-inferiority margin. |
| Discuss and agree choice of non-inferiority margin with regulators before conducting the trial and where regulatory disease guidelines describe a specific non-inferiority margin, link this with a specific estimand. |
| Provide transparent, precise and comprehensive description of the estimand(s) and intercurrent events distribution when reporting results from clinical trials. |

**Acknowledgements**

The authors thank Bryan Goldman (Novo Nordisk) for discussions related to the STEP trials and Claudia Hemmelmann (Sandoz) and Chien-Ju Lin (Roche) for helpful discussions on the content of this manuscript.

**Conflict of interest disclosure**

The following potential competing interests are declared:

H.L., T.M., M.M., S.R. and D.W. are employed by their affiliated pharmaceutical companies as declared above that are involved in non-inferiority clinical trials. O.N.K. is an independent consultant to the pharmaceutical industry and has previously held positions with GSK. H.L., S.R., M.M., T.M. and D.W. hold shares in their respective companies.

# Supporting information to "Estimands and the Choice of Non-Inferiority Margin under ICH E9(R1)" by Mütze et al.

## Table of Contents

# S1  Simulation setup

*Table S1 Relative mean percentage change from baseline for different study visits.*

| Visit | Relative Mean %-change Placebo | Relative Mean %-change Reference |
|---|---|---|
| 0 | 0 | 0 |
| 4 | -1.12 | -2.32 |
| 8 | -1.70 | -4.07 |
| 12 | -2.20 | -6.05 |
| 16 | -2.51 | -7.84 |
| 20 | -2.86 | -9.71 |
| 28 | -2.89 | -12.04 |
| 36 | -3.04 | -13.86 |
| 44 | -3.31 | -15.21 |
| 52 | -3.30 | -16.14 |
| 60 | -3.22 | -16.53 |
| 68 | -3.02 | -16.8 |

The variance was assumed to be 95 at each visit, with an autoregressive covariance matrix structure and a correlation of 0.8 between neighboring visits.

# S2   Example 1

## S2.1   STEP trials

*Table 2* Estimands in the STEP trials. For all estimands, the population-level summary is difference in means (between treatment conditions).

| Trial | Target population | Treatment conditions | Endpoint | Intercurrent events and strategies |
|---|---|---|---|---|
| STEP 0 | Adults **without diabetes**, and with a body-mass index (BMI) of 30 kg/m² or more that was not of endocrine aetiology (e.g., Cushing's syndrome). | Once-daily (OD) subcutaneous (s.c.) 0.05 mg, or 0.1 mg or 0.2 mg or 0.3 mg or 0.4 mg semaglutide or OD 3.0 mg liraglutide or placebo all as adjunct to nutritional and physical counselling<br>1) regardless of treatment discontinuation<br>2) had all remained on treatment | Percentage change in bodyweight from baseline to week 52 | • Treatment discontinuation<br><br>Estimand 1: Handled by treatment policy strategy<br><br>Estimand 2: Handled by a hypothetical strategy |
| STEP 1 | Adults with a body-mass index of 30 or greater (≥27 in persons with ≥1 weight-related coexisting condition), **who did not have diabetes** | Once-weekly (OW) s.c. semaglutide or placebo both as adjunct to diet and lifestyle intervention<br>1) regardless of treatment discontinuation or initiation of rescue interventions<br>2) had treatment not been discontinued and had rescue medication not been available | Percentage change in bodyweight from baseline to week 68 | • Treatment discontinuation<br>• Use of rescue intervention<br><br>Estimand 1: Both handled by treatment policy strategy<br><br>Estimand 2: Both handled by a hypothetical strategy |
| STEP 2 | Adults with a body-mass index of at least 27 kg/m² and glycated haemoglobin 7–10% (53–86 mmol/mol) **who had been diagnosed with type 2 diabetes** at least 180 days | OW s.c. semaglutide 2.4 mg, OW s.c. semaglutide 1.0 mg or placebo all as adjunct to diet and lifestyle intervention<br>1) regardless of treatment discontinuation or initiation of rescue interventions | Percentage change in bodyweight from baseline to week 68 | • Treatment discontinuation<br>• Use of rescue intervention<br><br>Estimand 1: Both handled by treatment policy strategy |

| Trial | Target population | Treatment conditions | Endpoint | Intercurrent events and strategies |
|---|---|---|---|---|
| | | 2) had treatment not been discontinued and had rescue medication not been available | | Estimand 2: Both handled by a hypothetical strategy |
| STEP 3 | Adults **without diabetes** and with either overweight (body mass index>=27) plus at least 1 comorbidity or obesity (body mass index>=30). | OW s.c. semaglutide 2.4 mg or placebo both as adjunct to diet and lifestyle intervention 1) regardless of treatment discontinuation or initiation of rescue interventions 2) had treatment not been discontinued and had rescue medication not been available | Percentage change in bodyweight from baseline to week 68 | • Treatment discontinuation • Use of rescue intervention  Estimand 1: Both handled by treatment policy strategy  Estimand 2: Both handled by a hypothetical strategy |
| STEP 4 | Adults with body mass index of at least 30 (or >=27 with >=1 weight-related comorbidity) and **without diabetes**. who were able to titrate semaglutide to 2.4mg/wk within 20 weeks. | OW s.c. semaglutide 2.4 mg or placebo both as adjunct to diet and lifestyle intervention 1) regardless of treatment discontinuation or initiation of rescue interventions 2) had treatment not been discontinued and had rescue medication not been available | Percentage change in bodyweight from baseline (week 20) to week 68 | • Treatment discontinuation • Use of rescue intervention  Estimand 1: Both handled by treatment policy strategy  Estimand 2: Both handled by a hypothetical strategy |
| STEP 5 | Adults with overweight (BMI>= 27 with at least one weight-related comorbidity) or obesity (BMI>=30) | OW s.c. semaglutide 2.4 mg or placebo both as adjunct to diet and lifestyle intervention 1) regardless of treatment discontinuation or initiation of rescue interventions 2) had treatment not been discontinued and had rescue medication not been available | Percentage change in bodyweight from baseline to week 52 | • Treatment discontinuation • Use of rescue intervention  Estimand 1: Both handled by treatment policy strategy  Estimand 2: Both handled by a hypothetical strategy |
| STEP 6 | Adults from Japana or South Korea with overweight (BMI>= 27 | OW s.c. semaglutide 2.4 mg, OW s.c. semaglutide 1.7 mg or placebo all as | Percentage change in bodyweight from | • Treatment discontinuation |

| Trial | Target population | Treatment conditions | Endpoint | Intercurrent events and strategies |
|---|---|---|---|---|
| | with at least one weight-related comorbidity) or obesity (BMI>=35) | adjunct to diet and lifestyle intervention 1) regardless of treatment discontinuation or initiation of rescue interventions 2) had treatment not been discontinued and had rescue medication not been available | baseline to week 68 | • Use of rescue intervention<br><br>Estimand 1: Both handled by treatment policy strategy<br><br>Estimand 2: Both handled by a hypothetical strategy |
| STEP 7 | Adults from China, Hong Kong, Brazil or South Korea with overweight or obesity, **with or without type 2 diabetes** | OW s.c. semaglutide 2.4 mg or placebo both as adjunct to diet and lifestyle intervention 1) regardless of treatment discontinuation or initiation of rescue interventions 2) had treatment not been discontinued and had rescue medication not been available | Percentage change in bodyweight from baseline to week 44 | • Treatment discontinuation<br>• Use of rescue intervention<br><br>Estimand 1: Both handled by treatment policy strategy<br><br>Estimand 2: Both handled by a hypothetical strategy |
| STEP 8 | US adults BMI >= 30 or >=27 with weight-related comorbidity, **without diabetes** | OW s.c. semaglutide 2.4 mg, OD s.c. liraglutide 3.0 mg or placebo all as adjunct to diet and lifestyle intervention 1) regardless of treatment discontinuation or initiation of rescue interventions 2) had treatment not been discontinued and had rescue medication not been available | Percentage change in bodyweight from baseline to week 68 | • Treatment discontinuation<br>• Use of rescue intervention<br><br>Estimand 1: Both handled by treatment policy strategy<br><br>Estimand 2: Both handled by a hypothetical strategy |
| STEP 9 | Adults with obesity and a clinical and radiologic diagnosis of moderate knee osteoarthritis with at least moderate pain | OW s.c. semaglutide 2.4 mg or placebo both as adjunct to diet and lifestyle intervention 1) regardless of treatment discontinuation or | Percentage change in bodyweight from baseline to week 68 | • Regardless of adherence to the assigned regimen<br>• Use of other interventions (anti-obesity therapies |

| Trial | Target population | Treatment conditions | Endpoint | Intercurrent events and strategies |
|---|---|---|---|---|
| | | initiation of rescue interventions<br>2) had treatment not been discontinued and had rescue medication not been available | | or knee osteoarthritis interventions)<br>• Adherence to pain-medication washout |
| STEP 10 | Individuals aged 18 years or older with a BMI of 30 kg/m² or higher and **prediabetes** according to UK National Institute for Health and Care Excellence criteria | OW s.c. semaglutide 2.4 mg or placebo both as adjunct to diet and lifestyle intervention<br>1) regardless of treatment discontinuation or initiation of rescue interventions<br>2) had treatment not been discontinued and had rescue medication not been available | Percentage change in bodyweight from baseline to week 52 | • Treatment discontinuation<br>• Dose reduction of randomised treatment<br>• Initiation of other glucose-lowering medication<br>• Weight management therapies |
| STEP TEENS | Adolescents (12 to <18 years of age) with obesity (a body-mass index in the 95th percentile or higher) or with overweight (a BMI in the 85th percentile or higher) and at least one weight-related coexisting condition, and who were able to follow a 12-week lifestyle intervention run-in period | OW s.c. semaglutide 2.4 mg or placebo both as adjunct to diet and lifestyle intervention<br>1) regardless of treatment discontinuation or initiation of rescue interventions<br>2) had treatment not been discontinued and had rescue medication not been available | Percentage change in bodyweight from baseline to week 68 | • Treatment discontinuation<br>• Use of rescue intervention<br><br>Estimand 1: Both handled by treatment policy strategy<br><br>Estimand 2: Both handled by a hypothetical strategy |

*Table 3* Estimated treatment differences (ETD) with 95% confidence intervals, number of participants randomized and number of intercurrent events by treatment and trial.

| Trial | | Semaglutide | Placebo |
|---|---|---:|---:|
| STEP 1 | Randomised | 1306 | 655 |
| | Treatment discontinuation | 223 | 147 |
| | Anti-obesity intervention | 7 | 13 |
| | ETD [95% CI] (TP) | -12.44% [-13.37;-11.51] | |
| | ETD [95% CI] (HYP) | -14.4% [-15.3;-13.5] | |
| STEP 2 | Randomised | 404 | 403 |
| | Treatment discontinuation | 47 | 56 |
| | Anti-obesity intervention | 4 | 13 |
| | ETD [95% CI] (TP) | -6.21% [-7.28;-5.15] | |
| | ETD [95% CI] (HYP) | -7.6% [-8.6;-6.6] | |
| STEP 3 | Randomised | 407 | 204 |
| | Treatment discontinuation | 31 | 13 |
| | Anti-obesity intervention | Unknown | Unknown |
| | ETD [95% CI] (TP) | -10.3% [-12;-8.6] | |
| | ETD [95% CI] (HYP) | -12.7% [-14.3;-11] | |
| STEP 4 | Randomised | 535 | 268 |
| | Treatment discontinuation | 31 | 31 |
| | Anti-obesity intervention | 0 | 1 |
| | ETD [95% CI] (TP) | -14.8% [-16;-13.5] | |
| | ETD [95% CI] (HYP) | -15.3% [-16.5;-14.1] | |
| STEP 5 | Randomised | 152 | 152 |
| | Treatment discontinuation | 20 | 41 |
| | Anti-obesity intervention | Unknown | Unknown |
| | ETD [95% CI] (TP) | -12.6% [-14.5;-10.7] | |
| | ETD [95% CI] (HYP) | -14.3% [-16.1;-12.4] | |
| STEP 6 | Randomised | 199 | 101 |
| | Treatment discontinuation | 13 | 3 |
| | Anti-obesity intervention | Unknown | Unknown |
| | ETD [95% CI] (TP) | -11.1% [-12.9;-9.2] | |
| | ETD [95% CI] (HYP) | -11.35% [-13.2;-9.5] | |
| STEP 7 | Randomised | 249 | 126 |
| | Treatment discontinuation | 18 | 16 |
| | Anti-obesity intervention | Unknown | Unknown |
| | ETD [95% CI] (TP) | -8.5% [-10.2;-6.8] | |

| Trial | | Semaglutide | Placebo |
|---|---|---|---|
| | ETD [95% CI] (HYP) | -9.9% [-11.3;-8.4] | |
| | | | |
| STEP 8 | Randomised | 126 | 85 |
| | Treatment discontinuation | 17 | 15 |
| | Anti-obesity intervention | 1 | 3 |
| | ETD [95% CI] (TP) | -13.9% [-16.7;-11] | |
| | ETD [95% CI] (HYP) | -15.3% [-17.9;-12.7] | |
| | | | |
| STEP 9 | Randomised | 271 | 136 |
| | Treatment discontinuation | 34 | 29 |
| | Anti-obesity intervention | Unknown | Unknown |
| | ETD [95% CI] (TP) | -10.5% [-12.3;-8.6] | |
| | ETD [95% CI] (HYP) | -12.1% [-13.8; -10.5] | |
| | | | |
| STEP 10 | Randomised | 138 | 69 |
| | Treatment discontinuation | 18 | 8 |
| | Anti-obesity intervention | Unknown | Unknown |
| | ETD [95% CI] (TP) | -11.2% [-13;-9.4] | |
| | ETD [95% CI] (HYP) | -13.3% [-15.2;-11.4] | |

## S2.2 Meta-analysis

### S2.2.1 R Code for meta-analysis

The following code was used to perform the meta-analysis in Section 4 of the main manuscript:

```
gMAP(cbind(trtDiff, trtDiffSE) ~ 1|trial, data = trtPolData, tau.dist = "HalfNormal", tau.prior = 5
```

The dataset included the following data:

| Trial | Estimand | Treatment difference | Standard error of treatment difference |
|---|---|---|---|
| STEP 1 | Treatment Policy | -12.44 | 0.474499 |
| STEP 1 | Hypothetical | -14.4 | 0.459192 |
| STEP 2 | Treatment Policy | -6.21 | 0.545928 |
| STEP 2 | Hypothetical | -7.6 | 0.510213 |
| STEP 3 | Treatment Policy | -10.3 | 0.867363 |
| STEP 3 | Hypothetical | -12.7 | 0.816342 |
| STEP 5 | Treatment Policy | -12.6 | 0.969406 |
| STEP 5 | Hypothetical | -14.3 | 0.918384 |
| STEP 6 | Treatment Policy | -11.1 | 0.918384 |
| STEP 6 | Hypothetical | -11.35 | 0.943895 |
| STEP 8 | Treatment Policy | -13.9 | 1.428598 |
| STEP 8 | Hypothetical | -15.3 | 1.326555 |
| STEP 9 | Treatment Policy | -10.5 | 0.918384 |
| STEP 9 | Hypothetical | -12.1 | 0.867363 |
| STEP 10 | Treatment Policy | -11.2 | 0.918384 |
| STEP 10 | Hypothetical | -13.3 | 0.969406 |

### S2.2.2 Sensitivity analysis

Meta-analysis results excluding STEP 2

| Handling strategy for discontinuation and other anti-obesity intervention | Pooled % mean treatment-effect and 95% credible interval |
|---|---|
| Treatment policy | -11.7 (-12.7, -10.5) |
| Hypothetical | -13.4 (-14.6, -12.0) |

# S3 Example 2

The following presents information extracted from each of the SCALE studies defining each estimand attribute relating directly to information held within the published article or the article's supplementary information. The first table for the SCALE maintenance, SCALE obesity & pre-diabetes and SCALE sleep apnea studies presents four attributes: population, endpoint/variable, treatment and population-level summary.

The second table presents the information extracted with regards to the analyses conducted aligned with the endpoint planned for our new study and any sensitivity analyses.

The third table presents information taken from the disposition table for each study based on withdrawals. This includes our inference for which events are considered as intercurrent events and which are considered as study withdrawals.

The fourth study is the only one that specified the estimand fully. Information on the analyses performed are included along with the number of patients that had the intercurrent event and withdrew from the study.

## S3.1 SCALE Maintenance study

*Table 4 Estimand information for the SCALE Maintenance study.*

| Attribute | Description | Quote / Justification |
|---|---|---|
| **Population** | Overweight or obese adults (BMI ≥30, or ≥27 with comorbidity) who first achieved ≥5% weight loss via low-calorie diet run-in. | "Obese/overweight participants (≥18 years, body mass index ≥30 kg/m$^2$ or ≥27 kg/m$^2$ with comorbidities) who lost ≥5% of initial weight during a LCD run-in were randomly assigned to liraglutide 3.0mg per day or placebo (subcutaneous administration) for 56 weeks." |
| **Endpoint** | Percentage change in fasting body weight between baseline and week 56. | "Three co-primary end points were tested hierarchically at week 56: (1) mean percentage change in fasting body weight from randomization; (2) the proportion of individuals that maintained the ≥5% reduction in fasting body weight achieved during LCD run-in; and (3) the proportion that lost ≥5% of fasting body weight after randomization."<br><br>Results for the primary endpoint present: Change from randomization to week 56. |
| **Treatment** | Liraglutide 3.0 mg once daily SC + diet and exercise counselling vs. placebo + diet and exercise counselling. | "To facilitate dietary adherence, participants met face-to-face, every other week with a nutritionist and had telephone calls on alternate weeks. They were encouraged to exercise regularly (recommended 150 min per week of brisk walking) and were provided with pedometers. As soon as individuals lost ≥5% of screening body weight, they were randomly assigned 1:1 to receive once-daily liraglutide 3.0mg (n=212) or placebo (n=210). Randomization was performed centrally using a telephone- or web-based system. |

|  |  | Participants were stratified according to comorbidity status and BMI. Treatment allocation was blinded to participants, investigators, and sponsors throughout the trial. Liraglutide (6.0mg per ml) and placebo were provided in modified FlexPen devices and administered subcutaneously. Dosing was escalated weekly to reach 3.0 mg. Participants were prescribed a 500 kcal/day deficit diet with specified macronutrient intake. Face-to-face lifestyle counselling visits were provided 17 times over 56 weeks; medical monitoring occurred on the same schedule." |
|---|---|---|
| **Population-level summary** | Difference in means. | Estimated treatment difference for liraglutide versus placebo (95% CI), P-value is reported in Table 3. |

*Table 5 Analysis information for the SCALE Maintenance study. Full analysis set: comprising all randomized individuals exposed to trial drug with at least one post-randomization weight assessment.*

| Analysis | Definition / Description | Details in the paper/SI |
|---|---|---|
| **Primary analysis** | Continuous efficacy endpoints: change from randomization to week 56 in fasting body weight. | <ul><li>Conducted in full analysis set (FAS: all randomized individuals exposed to trial drug with at least one post-randomization weight assessment).</li><li>ANCOVA for continuous endpoints.</li><li>Included treatment, sex, country and comorbidity stratification as fixed effects, and randomization value as a covariate.</li><li>Missing values imputed using last observation carried forward (on drug).</li></ul> |

| Per-protocol (completer) analysis | Included all participants from the full analysis set who did not significantly violate inclusion, exclusion or randomization criteria, who had a valid assessment at week 56, where week 56 was at least 52 weeks (365 days) after first drug date, but who were allowed to be off drug for a total of 4 weeks during the trial, but at most 2 consecutive weeks. No imputation was performed on the data. Body weight was measured in the fasting state. | • Completers analysis.<br>• No imputation performed. |
|---|---|---|
| Repeated measures analysis | Linear mixed-effect model (MMRM) was done using the full analysis set with no imputation. Body weight was measured in the fasting state. | • FAS population.<br>• Fixed effects: treatment, sex, country, comorbidity stratification; covariate: baseline value; includes treatment-by-visit interaction.<br>• Unstructured covariance for residuals within subjects. |
| Analysis including fasting and non-fasting measurements | Including **off-drug** and non-fasting weight measurements. | • FAS population.<br>• ANCOVA model as per primary analysis.<br>• Missing values imputed by last observation carried forward. |
| Secondary analysis | Change from randomization to week 64 (including 12-week off-drug follow-up). | • ANCOVA for continuous endpoints. Same fixed effects and covariates as primary analysis.<br><br>• Changes from randomization to week 56 or 68 are observed means (s.d.). ETDs are from an analysis of covariance and OR are from a logistic regression analysis, all using the full analysis set and with the last observation carried forward, except for percentage weight-loss data at week 68, which was without last observation carried forward. |

*Table 6* Information on potential intercurrent events taken from the disposition table. Total withdrawals: Liraglutide 3.0 mg = 53, Placebo = 64.

| Withdrawal reasons | Liraglutide 3.0 mg (N randomised = 212) | Placebo (N randomised = 210) | Intercurrent event | Notes |
|---|---|---|---|---|
| Discontinuation due to adverse events | 18 | 18 | Yes | A clear post-baseline event affecting either the interpretation or the existence of the measurements associated with the clinical question of interest |
| Discontinuation due to lack of efficacy | 0 | 2 | Yes | A clear post-baseline event affecting either the interpretation or the existence of the measurements associated with the clinical question of interest |
| Non-compliance | 8 | 5 | No | Administrative/procedural withdrawals. Considered as missing data. |
| Withdrawal criteria | 17 | 24 | No | Administrative/procedural withdrawals. Considered as missing data. |
| Other reasons | 10 | 15 | No | Administrative/procedural withdrawals. Considered as missing data. |

## S3.2 SCALE Obesity & Prediabetes

*Table 7* Estimand information for the SCALE Obesity & Prediabetes trial.

| Attribute | Description | Quote / Justification |
|---|---|---|
| Population | Overweight/obese adults (BMI ≥30, or ≥27 with comorbidity), without type 2 diabetes. | "We conducted a 56-week, double-blind trial involving 3731 patients who did not have type 2 diabetes and who had a body-mass index (BMI; the weight in kilograms divided by the square of the height in meters) of at least 30 or a BMI of at least 27 if they had treated or untreated dyslipidemia or hypertension." |
| Endpoint | Percentage change in body weight between baseline and week 56. | "The three prespecified coprimary end points, assessed at week 56, were weight change from baseline, the proportion of patients who lost at least 5% of their baseline body weight, and the proportion of patients who lost more than 10% of their baseline body weight."<br><br>Results present: Changes in Coprimary End Points and Cardiometabolic Risk Factors between Baseline and Week 56. |

| Treatment | Liraglutide 3.0 mg daily SC + lifestyle counselling vs. placebo + lifestyle counselling. | "Eligible patients were randomly assigned, in a 2:1 ratio, to receive once-daily subcutaneous injections of liraglutide, starting at a dose of 0.6 mg with weekly 0.6-mg increments to 3.0 mg, or placebo; both groups received counseling on lifestyle modification (Fig. S1 in the Supplementary Appendix)." |
|---|---|---|
| Population-level summary | Difference in means. | "After 56 weeks, patients in the liraglutide group had lost a mean (±SD) of 8.0±6.7% (8.4±7.3 kg) of their body weight, whereas patients in the placebo group had lost a mean of 2.6±5.7% (2.8±6.5 kg) of their body weight (Table 2)." |

*Table 8* Analysis information for the SCALE Obesity & Prediabetes trial. Full analysis set included all patients who underwent randomization and received at least one dose of a study drug and had at least one assessment after baseline.

| Analysis | Definition / Population | Details in the paper/SI |
|---|---|---|
| **Primary analysis** | FAS: all patients who underwent randomization and received at least one dose of a study drug and had at least one assessment after baseline | Missing values were imputed with the use of the last-observation-carried-forward method for measurements made after baseline. For weight, only fasting measurements were used. The three coprimary end points were analyzed in hierarchical order. An analysis of covariance model was used to analyze mean changes in continuous end points. The model included treatment, country, sex, BMI stratification, status with respect to prediabetes at screening, and interaction between BMI strata and prediabetes status as fixed effects, with the baseline value of the relevant variable as a covariate. |
| **Completers analysis** | FAS subjects with valid, non-imputed week 56 measurement | The primary analysis will be repeated based on completers (week 56) in FAS with a valid nonimputed measurement at week 56 |
| **Repeated measures** | All randomized subjects | The primary analysis will be repeated including all randomised subjects allowing for baseline observation carried forward for subjects without a post baseline measurement. |
| **Baseline observation carried forward** | FAS subjects | Primary analysis repeated on FAS including fasting and non-fasting weight measurements, off-drug measurements, and follow-up weight measurements after week 56 (visit 17x). |
| **All available measurements** | FAS subjects | The primary analysis will be based on subjects in FAS, including the fasting and non-fasting weight measurements, off drug measurements, and the follow-up measurements after 56 weeks of randomisation (visit 17x) and weight measurements following rescue medication |

| Multiple imputation | FAS subjects with missing post-baseline fasting body weight imputed | - The primary analysis will be repeated based on subjects in FAS, but by imputing missing postbaseline observations of the primary endpoint at visits where fasting body weight is scheduled to be measured using a multiple imputation procedure. Missing values in the two treatment arms will be imputed based on the response profile of completers in the placebo arm. The imputation will be performed according to the following steps:<br>  o Intermittent missing values of fasting body weight are imputed separately within each treatment group based on observed data using a Markov Chain Monte Carlo method were 100 copies of the dataset will be generated<br>  o For each of the 100 copies, missing values at visits where fasting body weight is scheduled to be measured in a fasting state will be imputed sequentially using an ANCOVA model including treatment, country, gender, BMI stratification factor, pre-diabetes status at screening and an<br>  o interaction between BMI strata and pre-diabetes status at screening as fixed factors and the fasting weight measurements at the previous scheduled visits as covariates. The sequential imputation means that missing values at the first scheduled visit post baseline will be imputed based on the above mentioned model where the baseline value is included as a covariate. Similarly, missing values at the second scheduled visit post baseline will be imputed based in the above mentioned model where both the baseline value and the value at the first scheduled post-baseline visit, which might be imputed, are included as covariates. Missing values at each scheduled post baseline visit will be imputed in this manner until values at 56 weeks are obtained.<br>  o For each of the 100 copies, the relative change from baseline in fasting body weight will be derived and the same model as used in the primary analysis will be used to analyse the values at 56 weeks.<br>  o The estimated treatment effects, as well as the standard errors, obtained from analysing each of the 100 copies in the previous step will be pooled into one estimate from which the 95% confidence interval and p-value are calculated. |

|  | • All non-imputed measurements of the primary endpoint on post baseline visits where fasting body weight is scheduled to be measured will be analysed using a repeated measurements model based on subjects in FAS. The model will include the same fixed factors and covariates as stated for the primary analysis but these will be nested within visit in this model, i.e. the effect of each factor and covariate will be assumed to depend on visit. An unstructured covariance structure for the residuals within subjects will be assumed and residuals from different subjects will be assumed independent. From this model the treatment effect at 56 weeks will be estimated. |
|---|---|

*Table 9* Information on potential intercurrent events taken from the disposition table.

| Withdrawal reasons | Liraglutide 3.0 mg (N randomised = 2487) | Placebo (N randomised = 1244) | Intercurrent event | Notes |
|---|---|---|---|---|
| **Discontinuation due to adverse events** | 238 | 45 | Yes | Clear post-baseline events affecting either the interpretation or existence of measurements associated with the clinical question of interest |
| **Discontinuation due to ineffective therapy** | 23 | 36 | Yes | Clear post-baseline events affecting either the interpretation or existence of measurements associated with the clinical question of interest |
| **Non-compliance** | 65 | 38 | No | Administrative/procedural withdrawals. Considered as missing data. |
| **Withdrawal criteria (including AE withdrawals)** | 294 | 261 | No | Administrative/procedural withdrawals. Considered as missing data. |
| **Other reasons** | 78 | 63 | No | Administrative/procedural withdrawals. Considered as missing data. |

**Total withdrawals:** Liraglutide 3.0 mg = 698, Placebo = 443
**Withdrew but attended follow-up visit at week 56:** Liraglutide 3.0 mg = 202, Placebo = 111
**Notes:** Patients are excluded due to no exposure to treatment or no post-baseline values.

## S3.3  SCALE Sleep Apnea

| Attribute | Description | Quote / Justification |
|---|---|---|
| **Population** | Non-diabetic participants with moderate to severe sleep apnea who are unwilling or unable to use CPAP therapy, aged 18–64 years, with stable body weight (<5% change during previous 3 months) and BMI ≥30 kg/m². | "In brief, eligible individuals were men and women aged 18–64 years with a stable body weight (<5% change during the previous 3 months) and body mass index (BMI) of ⩾30 kg m−2. Eligible individuals had to be diagnosed with moderate (apnea–hypopnea index (AHI) 15.0–29.9 events h−1) or severe (AHI ⩾30.0 events h−1) OSA and be unable or unwilling to use CPAP therapy." |
| **Endpoint** | Percentage change in body weight between baseline and week 32. | " The primary efficacy end point was change in AHI (using the American Academy of Sleep Medicine's recommended 2007 definition, with hypopnea scoring requiring ⩾30% reduction in nasal pressure signal excursions from baseline and ⩾ 4% desaturation from pre-event baseline)29 from baseline to week 32. Key secondary efficacy end points included changes from baseline to week 32 in OSA severity category, blood oxygen saturation parameters (lowest oxygen saturation, percentage of time with oxygen saturation o90% and oxygen desaturation ⩾4% index), sleep architecture parameters (total sleep time, wake time after sleep onset, proportion of sleep spent in supine position and sleep stage distribution), body weight-related parameters (fasting body weight, proportion of participants losing ⩾5% or 410% of baseline fasting body weight, BMI, waist and neck circumference), glycemic parameters (HbA1c and fasting plasma glucose), vital signs (systolic blood pressure (SBP) and diastolic blood pressure, pulse), fasting lipids (high-density lipoprotein, low-density lipoprotein, very-low-density lipoprotein, and total cholesterol and triglycerides), cardiovascular biomarkers (high-sensitivity C-reactive protein and urinary albumin: creatinine ratio), daytime sleepiness (Epworth Sleepiness Scale) and self-reported quality of life (Functional Outcomes of Sleep Questionnaire (FOSQ) and 36-item Short-Form (SF-36) health status survey)."<br><br>Results present: Mean body weight decreased continuously in the liraglutide 3.0 mg group over the 32 weeks. |

| Treatment | Liraglutide 3.0 mg/day SC as adjunct to diet (500 kcal/day deficit) and exercise vs. placebo as adjunct to diet (500 kcal/day deficit) and exercise. | "Once-daily subcutaneous liraglutide or placebo. To reduce the likelihood of gastrointestinal symptoms, liraglutide was started at 0.6 mg/day and escalated in weekly 0.6-mg increments to 3.0 mg (week 4). The 3.0 mg dose was maintained for another 28 weeks. A placebo dose-volume equivalent was used to maintain blinding. Participants, investigators, and sponsor were blinded. All participants received counselling on diet and physical activity approximately every 4 weeks during treatment." |
|---|---|---|
| Population-level summary | Difference in mean. | "At week 32, the mean weight loss was significantly greater with liraglutide than with placebo (−5.7 ± 0.4% vs −1.6 ± 0.3%, estimated treatment difference: −4.2%, 95% CI −5.2 to −3.1%, P≤0.0001)." |

Information on analyses:

| Analysis | Definition / Population | Details in the paper/SI |
|---|---|---|
| Primary analysis | FAS: All patients randomized | Continuous end points were analyzed using a pre-specified analysis of covariance model with treatment, gender and country as fixed effects and baseline BMI, age and value at baseline as covariates. |

Information on potential intercurrent events taken from the disposition table:

| Withdrawal reasons | Liraglutide 3.0 mg (N randomised = 180) | Placebo (N randomised = 179) | Intercurrent event | Notes |
|---|---|---|---|---|
| **Discontinuation due to adverse events** | 20 | 6 | Yes | Clear post-baseline events affecting either the interpretation or existence of measurements associated with the clinical question of interest |
| **Discontinuation due to ineffective therapy** | 2 | 1 | Yes | Clear post-baseline events affecting either the interpretation or existence of measurements |

|  |  |  |  |  | associated with the clinical question of interest |
| --- | --- | --- | --- | --- | --- |
| **Non-compliance** | 8 | 5 |  | No | Administrative/procedural withdrawals. Considered as missing data. |
| **Withdrawal criteria (including 1 AE)** | 14 | 20 |  | No | Administrative/procedural withdrawals. |
| **Other reasons** | 2 | 5 |  | No | Administrative/procedural withdrawals. |

**Total withdrawals:** Liraglutide 3.0 mg = 46, Placebo = 37
**Notes:** For non-compliance, withdrawal criteria and other reasons includes incorrect handling of trial product, non-compliance to the visit schedule and dietary advice, incomplete questionnaires.

## S3.4 SCALE Insulin Trial

| Attribute | Primary / Treatment Policy | Secondary / Trial Product | Quote / Justification |
| --- | --- | --- | --- |
| **Population** | Adult patients aged ≥18 years with BMI ≥27 kg/m$^2$, stable body weight (≤5 kg change within 90 days before screening), diagnosed with type 2 diabetes (HbA1c 6.0–10%) and receiving stable treatment with any basal insulin (≥90 days) and ≤2 OADs. | Same as primary. | "Eligible individuals were aged ≥18 years with a BMI of ≥27 kg/m$^2$, stable body weight (≤5 kg self-reported change within 90 days), diagnosed with type 2 diabetes with HbA1c 6.0–10% at screening, and receiving stable treatment with any basal insulin (≥90 days; no requirement for minimum or maximum dose) and ≤2 OADs." |
| **Endpoint** | Percentage change in body weight from baseline to week 56. | Same as primary. | "Coprimary end points were change in body weight (percentage) from baseline to week 56 and proportion of individuals losing ≥5% of baseline body weight at week 56." |
| **Treatment** | Liraglutide 3.0 mg/day SC + intensive behavioural therapy (IBT) vs. placebo + IBT. | Same as primary. | "A total of 396 individuals were randomized centrally using an interactive voice/web response system to receive either liraglutide 3.0 mg or placebo (1:1) as adjunct to IBT." |
| **Population-level summary** | Difference in mean; treatment policy estimand (intention-to-treat). | Difference in mean; trial product estimand (if-all-adhered). | "For the treatment policy estimand, mean weight loss at 56 weeks was 25.8% with liraglutide 3.0 mg and 21.5% with placebo (ETD 4.3% [95% CI 5.5, 3.2]; P<0.0001)." |

| | | | For the trial product estimand, estimated mean weight change at 56 weeks was 26.4% for liraglutide 3.0 mg and 21.3% for placebo (ETD 5.1% [95% CI 6.3, 3.9]; P<0.0001)." |
|---|---|---|---|
| **Intercurrent events (handling strategy)** | Treatment discontinuation (Treatment policy) | Treatment discontinuation (Hypothetical) | **Estimation details for the primary estimand:** Statistical analysis is ANCOVA with jump-to-reference multiple imputation.<br><br>Missing values at week 56 were imputed from the placebo arm using a jump-to-reference multiple imputation approach based on 100 iterations of the data set.<br><br>**Estimation details for the secondary estimand:** The trial product estimand for the primary and confirmatory secondary endpoints was assessed using a mixed model for repeated measurements (MMRM). The MMRM for percentage weight change was done using assessments only from individuals who were taking the randomized treatment until end of trial or at first discontinuation of trial product. The MMRM was fitted using percentage weight change with factors treatment, sex and BMI, and baseline body weight as a covariate, all nested within visits. An unstructured covariance matrix was used to account for the variability for measurements within the same individual. |

### S3.4.1 Information on intercurrent events and missing data:

| Event | Liraglutide 3.0 mg (N randomised = 198) | Placebo (N randomised = 198) |
|---|---|---|
| **Intercurrent event: Discontinued treatment** | 26 | 26 |
| **Withdrew (missing data)** | 6 | 4 |

## S3.5  Summary of results from SCALE trials

*Table 10 Results presented from the SCALE program for the mean % change in body weight*

| Trial | Type of analysis[1] | N analysed (Liraglutide 3.0 mg s.c.) | N analysed (Placebo) | Liraglutide 3.0 mg s.c. Mean (SD)[2] | Placebo Mean (SD)* | Estimated treatment difference for liraglutide versus placebo (95% CI) |
|---|---|---|---|---|---|---|
| **SCALE Maintenance – without type 1 and type 2 diabetes** | | | | | | |
| | Primary analysis | 207 | 206 | -6.2 (7.3) | -0.2 (7.0) | -6.1 (-7.5, -4.6) |
| | Per-protocol (completer) analysis | 148 | 141 | -6.7 (7.7) | -0.1 (7.6) | -6.8 (-8.5, -5.0) |
| | Repeated measures analysis | 156 | 144 | -6.8 (7.8) | 0.0 (7.5) | -6.1 (-7.7, -4.6) |
| | Analysis with fasting and non-fasting observations | 212 | 208 | -5.2 (7.5) | 0.1 (7.3) | -5.4 (-6.8, -3.9) |
| | Secondary analysis (Week 68) | 159 | 144 | -4.1 (8.2) | 0.3 (7.7) | -4.2 (-6.0, -2.4) |
| **SCALE Obesity and pre-diabetes – without type 1 and type 2 diabetes** | | | | | | |
| | Primary analysis | 2437 | 1225 | –8.0 (6.7) | –2.6 (5.7) | –5.4 (–5.8 to –5.0) |
| | Completer population | 1781 | 798 | –9.2 | –3.5 | –5.7 (–6.3 to –5.1) |
| | Repeated measures | 2432 | 1220 | –8.5 | –2.7 | –5.8 (–6.3 to –5.3) |
| | Baseline observation carried forward | 2481 | 1239 | –7.8 | –2.6 | –5.3 (–5.7 to –4.8) |
| | All available measurements | 2437 | 1225 | –7.8 | –2.6 | –5.2 (–5.6 to –4.7) |
| | Multiple imputation | 2437 | 1225 | –8.3 | –2.7 | –5.5 (–6.0 to –5.0) |
| **SCALE Sleep Apnea – without type 1 and type 2 diabetes** | | | | | | |
| | Secondary analysis (Week 32) | 180 | 179 | –5.7 ± 0.4% | –1.6 ± 0.3% | –4.2 (–5.2 to –3.1) |
| **SCALE Insulin Trial – with type 2 diabetes and treated with basal insulin** | | | | | | |
| | Primary analysis (treatment policy) | 198 | 198 | -5.8 | -1.5 | –4.3 (–5.5 to –3.2) |
| | Secondary analysis (trial product) | 198 | 198 | -6.4 | -1.3 | –5.1 (–6.3 to –3.9) |

[1] All results presented at Week 56 unless otherwise specified.

[2] SD presented as ± or (SD) where available.